\documentclass[journal=jctc,manuscript=article]{achemso}
\usepackage{chemformula} 
\usepackage[T1]{fontenc} 
\usepackage{mhchem}

\def\lang{\langle}
\def\rang{\rangle}

\author{Yuhang Ai $^\dagger$}
\author{Ze-Wei Li $^\dagger$}
\author{Zhe-Bin Guan}
\author{Hong Jiang}
\email{jianghchem@pku.edu.cn}
\affiliation[Peking University]
{Beijing National Laboratory for Molecular Sciences, Institute of Theoretical and Computational Chemistry, College of Chemistry and Molecular Engineering, Peking University, Beijing 100871, China}

\title{Density Matrix Embedding Theory-Based Multi-Configurational Quantum Chemistry Approach to Lanthanide Single-Ion Magnets}

\begin{document}
\maketitle
\def\thefootnote{$\dagger$}\footnotetext{These authors contributed equally to this work}\def\thefootnote{\arabic{footnote}}

\date{\today}

\begin{abstract}
  Accurate and efficient theoretical descriptions of lanthanide systems based on \textit{ab initio} electronic structure theory remain highly challenging due to the complex interplay of strong electronic correlation and significant relativistic effects in $4f$ electrons. The composite multi-configurational quantum chemistry method, which combines the complete active space self-consistent field (CASSCF) approach with subsequent state interaction (SI) treatment of spin-orbit coupling (SOC), abbreviated as CASSI-SO, has emerged as the preferred method for \textit{ab initio} studies of lanthanide systems. However, its widespread application is hindered by its substantial computational cost. Building on the success of integrating density-matrix embedding theory (DMET) with CASSI-SO in our previous theoretical study of 3d single-ion magnets (SIMs) (Ai, Sun, and Jiang, \textit{J. Phys. Chem. Lett.} 2022, 13, 10627), we now extend the DMET+CASSI-SO approach to lanthanide SIM systems. We provide a detailed formulation of the regularized direct inversion of iterative subspace (R-DIIS) algorithm, which ensures obtaining physically correct restricted open-shell Hartree-Fock (ROHF) wavefunctions, a critical factor for the effectiveness of DMET. Additionally, we introduce the subspace R-DIIS (sR-DIIS) algorithm, which proves to be more efficient and robust for lanthanide systems. Using several representative lanthanide single-ion magnets (4f-SIMs) as test cases, we demonstrate the performance of these new algorithms and highlight the exceptional accuracy of the DMET+CASSI-SO approach. We anticipate that this enhanced DMET+CASSI-SO methodology will significantly advance large-scale theoretical investigations of complex lanthanide systems.
\end{abstract}

\section{1. Introduction}\label{sec:intro}

Electronic structure problems in lanthanide-based systems have long been considered challenging due to the strong correlation among 4f electrons, which is intricately intertwined with significant relativistic effects.\cite{Seijo2016, Dolg2015, Reid2016, Barandiaran2022} Among the various \textit{ab initio} theoretical methods developed for these systems\cite{Dolg2015, Reid2016, Barandiaran2022, Seijo2016, Ogasawara2008, Jia2016, Filatov1998, Frank1998, Ramanantoanina2015}, multi-configurational (MC) quantum chemistry methods \cite{Roos2016} have gained considerable popularity in recent years. This is due to 1) their exceptional ability to accurately address both static correlation (e.g., through complete active space (CAS) or restricted active space (RAS) self-consistent field (SCF) methods)\cite{Roos1980, Olsen1988, MaD2011} and dynamic correlation (e.g., via multi-reference configuration interaction, perturbation theory, or coupled cluster methods, MRCI/PT/CC) \cite{Roos2016, Lischka2018, Park2020}, and 2) their straightforward incorporation of relativistic effects at the two-component level within the framework of the state interaction spin-orbit coupling (SI-SO) approach \cite{Atanasov2015}. Consequently, the CASSI-SO method has been extensively applied to numerous $d$- or $f$-electron systems of interest in recent decades, including isolated molecules such as transition metal or lanthanide single-ion magnets (SIMs) \cite{Woodruff2013, Meng2016} and transition metal or lanthanide impurities in solids, such as lanthanide-doped luminescent materials \cite{Lucas2015, Seijo2016, Barandiaran2022}. These applications often yield satisfactory predictions, for example, of 4f-4f and 4f-5d excitation spectra \cite{Sun2023, Barandiaran2022}. However, the high theoretical accuracy achievable with these methods comes at the cost of significant computational complexity, making them often impractical when the active space or basis set becomes too large to handle. Previously, quantum embedding approaches \cite{Sun2016} such as density matrix embedding theory (DMET) \cite{Knizia2012, Wouters2016} have been proposed to significantly reduce the computational cost of multi-configurational quantum chemistry methods like CASSCF \cite{Roos1980} with minimal loss of accuracy. This strategy has been successfully demonstrated on typical correlated systems with 3d electrons, such as the iron porphyrin complex \cite{Pandharkar2019} and 3d-SIMs \cite{Ai2022}. Additionally, there have been attempts to use DMET+CAS and its derivatives to study molecular dissociation \cite{Pham2018, Hermes2019, Hermes2020}, vacancies in solids \cite{Mitra2021}, and molecules on surfaces \cite{Mitra2022}. Nevertheless, systems with lanthanide centers still present significant challenges for the DMET+CAS strategy. Obtaining a physically correct low-level wavefunction becomes more difficult due to the greater degeneracy of electronic states caused by unpaired 4f electrons, and the size of DMET impurities increases as the system grows. Furthermore, lanthanide centers are chemically distinct from 3d metals, as the more contracted 4f orbitals exhibit minimal interaction with the ligand field, and spin-orbit coupling (SOC) plays a much more prominent role. Therefore, whether DMET+CAS can be directly applied to lanthanide systems remains an open question that warrants further investigation.

In this work, we shall extend our previously proposed DMET+CASSI-SO scheme \cite{Ai2022} to systems with lanthanide centers, highlight the necessary adaptations, and investigate its performance on several 4f-SIMs. The structure of the paper is as the follows. We first review briefly the DMET+CASSI-SO theory and present a detailed formulation of the R-DIIS algorithm and its extension sR-DIIS. In Section 3, we demonstrate the performance and importance of using R-DIIS/sR-DIIS to obtain physically correct low-level wavefunction, and showcase the accuracy of DMET+CASSI-SO that can be achieved for typical 4f-SIMs with greatly reduced computational cost. The last section summarizes the main findings of this work.

\section{2. Methods and Computational Details}\label{sec:methods}

\subsection{2.1. DMET+CASSI-SO Theory}\label{sec:methods-dmet}

We first give a brief overview of the DMET+CASSI-SO approach, and more details can be found in our previous work \cite{Ai2022}. More comprehensive discussion about the methodological aspects of the DMET method can be found in refs. \citenum{Knizia2012, Wouters2016}. Within the framework of DMET \cite{Knizia2012},
using the Schmidt decomposition \cite{Peschel2012} of a low-level  Slater determinant wavefunction of the whole system
\begin{equation}
	|\Phi\rang=\sum_{I} \lambda_I|\Psi^{A}_I\rang\otimes |\Psi^{B_\mathrm{e}}_I\rang\otimes |\Phi^{B_\mathrm{c}}\rang=|\Psi^{
		A+B_\mathrm{e}}\rang\otimes |\Phi^{B_\mathrm{c}}\rang,
\end{equation}
we set up an impurity Hamiltonian in the DMET embedded cluster space $\mathcal{A}=A+B_\mathrm{e}$ encapsulating the cluster of interest defined by a set of localized orthogonal orbitals (e.g. for 4f-SIMs, those centered on the lanthanide ion), where $A$ denotes the cluster of interest, $B_\mathrm{e}$ and $B_\mathrm{c}$ denote bath and core orbitals, respectively. The union of the latter two forms the environment $B$, which is the complement of $A$. In practice, we obtain bath orbitals by diagonalizing the block of 1-particle spin-summed density matrix corresponding to the environmental orbitals and identifying the unentangled occupied (UO) orbitals (core orbitals), entangled orbitals (bath orbitals) and unentangled unoccupied orbitals (virtual orbitals) according to the eigenvalues $\zeta_i$ and a certain threshold $\varepsilon$ in form of $2-\varepsilon<\zeta_i$, $\varepsilon<\zeta_i<2-\varepsilon$ and $\zeta_i<\varepsilon$, respectively \cite{Wouters2016, Ai2022}. The larger $\varepsilon$ is, the smaller the size of the DMET impurity space would be. The DMET impurity Hamiltonian is then obtained based on projection to the embedded cluster space
\begin{equation}
	\hat H^\mathcal{A}=\hat H^\mathcal{A}_\mathrm{SR}+\hat H^\mathcal{A}_\mathrm{SOC}.
\end{equation}
$\hat H^\mathcal{A}_\mathrm{SR}$ is the scalar relativistic (SR) part
\begin{equation}
	\hat H^\mathcal{A}_\mathrm{SR}=\sum_{\mu\nu\in \mathcal{A}}\sum_{\sigma}h_{\mu\nu}^\mathcal{A}c_{\mu\sigma}^\dagger c_{\nu\sigma}+\sum_{\mu\nu\lambda\eta\in \mathcal{A}}\sum_{\sigma\sigma'}\lang \mu\nu|\lambda\eta\rang c_{\mu\sigma}^\dagger c_{\nu\sigma'}^\dagger c_{\eta\sigma'}c_{\lambda\sigma},
\end{equation}
with 
\begin{equation}\label{eq:h-A}
	\hat h^\mathcal{A} = -\frac{1}{2}\nabla^2-\sum_I\frac{Z_I}{|\mathbf{r}-\mathbf{R}_I|}+\sum_{b\in \mathrm{UO}}(2\hat J_b-\hat K_b).
\end{equation}
It should be noted that we give here the non-relativistic expression of $\hat H^\mathcal{A}_\mathrm{SR}$ to simplify the notation, and in practice, more sophisticated scalar relativistic single-particle Hamiltonians are usually used for systems with transition metal or lanthanide elements. $\hat H^\mathcal{A}_\mathrm{SOC}$ is the spin-dependent SOC term, which is approximated by the spin-orbit mean-field (SOMF) approach to the Pauli-Breit Hamiltonian in this work.\cite{Neese2005, Hess1996} Atomic units (a.u.) are used through the paper unless stated otherwise.

In the DMET+CASSI-SO method,\cite{Ai2022} $\hat{H}^\mathcal{A}_\mathrm{SR}$ is first solved by state-averaged (SA) CASSCF and the resulting configuration state functions $|\Psi_{N}^{SM}\rang$ (CSFs) are used to diagonalize $\hat{H}^\mathcal{A}$. The latter leads to a manifold of multiplet electronic state energies and wavefunctions, from which model spin Hamiltonian parameters \cite{Boca1999, Atanasov2015, Chibotaru2013, Malrieu2014} (e.g. zero-field splitting parameters) can be extracted for further theoretical analysis and simulations.\cite{Meng2016, Atanasov2015, Neese2002,Maganas2011,Maurice2009} Further consideration of dynamic correlation at the level of second order multi-configuration perturbation theory on top of SA-CASSCF, using either CASPT2 \cite{Andersson1992} or NEVPT2 \cite{Angeli2001}, is straightforward in both DMET or all electron treatment. Because computationally demanding high-level solvers like CASSCF work now in the much smaller DMET embedded cluster space $\mathcal{A}$, we are often able to observe significant speed boosts compared to all-electron calculations. Most importantly, as DMET is able to capture accurately the quantum entanglement between the cluster and the environment, the accuracy of DMET+CASSI-SO is comparable to the all-electron counterpart as well\cite{Ai2022}. An implementation \cite{liblan_code} of DMET+CASSI-SO can be found in our extension of the PySCF package \cite{Sun2018,Sun2020,Sun2015}.

\subsection{2.2. Acquiring Physically Correct Low-level Wavefunctions}\label{sec:methods-rdiis}
For open-shell systems, the restricted open-shell Hartree-Fock (ROHF) method is often used as the low-level solver to obtain the Slater determinant wavefunction that is used to build the DMET embedded Hamiltonian, especially when spin-pure orbitals are desirable, e.g. for subsequent SOC consideration \cite{Pandharkar2019,Mitra2021,Ai2022}.
In our previous work about DMET+CASSI-SO for 3d SIMs \cite{Ai2022}, we have observed that in typical 3d transition metal complexes there usually exist several ROHF solutions, and the accuracy of DMET+CASSI-SO as compared to its all-electron counterpart strongly depends on which solution is used. To be more specific, we found that the ROHF solution obtained by the default SCF settings often leads to significant discrepancies between DMET and all-electron results, which was found to be related to significant spin polarization in the ligand environment surrounding the transition metal ion. We proposed a new SCF technique named as R-DIIS that ensures obtaining the ROHF solution of lower energy with spin polarization dominantly localized on the transition metal ion, which, when used as the low-level solver, was found to be able to greatly improve the agreement between DMET and all-electron treatments. In this work, we will provide a more detailed formulation of the R-DIIS algorithm and present a further extension that is especially important for efficient treatment of lanthanide complexes.

As naive implementations of the SCF procedure often suffer from convergence problems, a lot of SCF accelerating techniques have been proposed in quantum chemistry \cite{Schlegel1991}, among which the direct inversion of iterative subspace (DIIS) method proposed by Pulay \cite{Pulay1982} is one of most widely used. In the original formulation of DIIS, the input density matrix $\mathbf{D}^{(n)}_{\rm in}$ of the $n$-th iteration is determined as a linear combination of output density matrices of previous $m$ iterations $\{\mathbf{D}^{(n-1)},\cdots,\mathbf{D}^{(n-m)}\}$ \cite{Pulay1982}
\begin{equation}
	\mathbf{D}^{(n)}_{\rm in}=\sum_{i=1}^{m} c_i\mathbf{D}^{(n-i)},
\end{equation}
and the coefficients $c_i$ are determined by the following optimization problem
\begin{equation}
	\min \left\Vert \sum_i c_i \mathbf{e}_i \right\Vert \quad s.t. \quad \sum_i c_i =1,
\end{equation}
where $\Vert\cdot\Vert$ denotes the $L_2$-norm, and $\mathbf{e}$ denotes the error vector defined as $\mathbf{e}\equiv \mathbf{SDF}-\mathbf{FDS}$, with $\mathbf{S}$ and $\mathbf{F}$ denoting the overlap and Fock matrices, respectively, which will be exactly zero when SCF converges \cite{Helgaker2000}.

Using DIIS can usually accelerate the SCF convergence significantly. However, if there exist multiple SCF solutions, DIIS can only help converging to the solution that is closest to the initial guess instead of the truly ground state solution or the desired solution with certain target properties (e.g. negligible spin polarization in the environment). The idea of R-DIIS is to block the unwanted solutions, and if possible, direct the SCF procedure to the targeted solution. To prevent converging to unwanted solutions, and using the fact that the diagonal elements of $\mathbf{e}$ are always zero by definition, we define a regularized error vector as
\begin{equation}
	\mathbf{e}'\equiv\mathbf{e}+\lambda \mathbf{I}R=\mathbf{SDF}-\mathbf{FDS}+\lambda \mathbf{I}R,
\end{equation}
where $R$ should be a quantity that vanishes as a necessary condition for the reaching of targeted SCF solution, and the parameter $\lambda$ is introduced to control the weight of the regularization term. As long as $R$ is not vanishing, the diagonal term $\lambda \mathbf{I} R$ would never get cancelled by $\mathbf{e}$ (all off-diagonal) and thus effectively blocks convergence to undesired solutions.

The specific form of $R$ is not unique, and can vary in different scenarios. For single-ion magnets with transition metal or lanthanide ions as magnetic centers, the physically correct low-level wavefunction should have spin-polarization dominantly localized on metal 3d/4f orbitals, and we proposed in our previous work \cite{Ai2022} to set $R$ as the environment spin-polarization entropy defined as
\begin{equation}
	\Delta S_\mathrm{E} = -2\mathrm{\,Tr}\left(\frac{\mathbf{D}_\mathrm{E}}{2} \ln \frac{\mathbf{D}_\mathrm{E}}{2}\right) +\mathrm{\,Tr}(\mathbf{D}^\alpha_\mathrm{E} \ln \mathbf{D}^\alpha_\mathrm{E})+\mathrm{\,Tr}(\mathbf{D}_\mathrm{E}^\beta \ln \mathbf{D}_\mathrm{E}^\beta).
\end{equation}
Here $\mathbf{D}_\mathrm{E}^{\sigma}$ ($\sigma=\alpha,\beta)$ refers to the spin-resolved truncated density matrix of the environment, and $\mathbf{D}_\mathrm{E}=\mathbf{D}_\mathrm{E}^{\alpha}+\mathbf{D}_\mathrm{E}^{\beta}$. $\Delta S_\mathrm{E}$ vanishes only if $\mathbf{D}^{\alpha}_{\rm E}=\mathbf{D}^{\beta}_{\rm E}$, which means no spin polarization in the environment and in turn all spin polarization on the metal center. 
We note in passing that the R-DIIS technique proposed here share some conceptual similarity with other SCF techniques seeking for multiple SCF solutions such as SCF meta-dynamics \cite{Thom2008}, the maximum overlap method \cite{Gilbert2008}, and the state-targeted energy projection \cite{CarterFenk2020}.

In our previous work in 3d SIMs \cite{Ai2022}, we found that R-DIIS can always help to find ROHF solutions of lower energy than that obtained from DIIS of the default setting with nearly vanishing $\Delta S_E$. On the other hand, R-DIIS has the disadvantage that it usually requires much larger number of iterations to converge than DIIS, which becomes more severe for lanthanide systems. Compared to 3d-SIMs, lanthanide complexes often have more unpaired electrons leading to greater degeneracy, and one often needs much more SCF cycles before R-DIIS SCF could converge, which means great cost if the SCF mixes all orbitals in each cycle. Another problem associated with the growing size of the system is that our desired solution space now also occupies a smaller portion of the entire solution space, as there are more molecular orbitals to mix around, and searching for them though a random-walk-like mechanism could be more difficult. To cut down the cost of R-DIIS and enlarge the portion of the desired solution space, we may restrict the R-DIIS routine to the DMET embedded cluster space computed from a loosely converged wavefunction by using the default DIIS, such that SCF facilitated by R-DIIS in this subspace only involves orbitals with large overlaps with the metal ion.
This strategy, termed as the subspace R-DIIS (sR-DIIS) henceforth, is based on the observation that the presence of multiple ROHF solutions can be attributed to different occupation of nearly degenerate orbitals around the Fermi level. A loosely-converged DIIS wavefunction should have most of unentangled occupied and virtual orbitals correct as indicated by the small variation in total energy near convergence. By carrying out R-DIIS in the much smaller embedded cluster space, singly-occupied orbitals with physically incorrect features acquired by original DIIS are rotated with other orbitals in the embedded cluster space such that they become mainly localized on the metal center. Combined with the unentangled occupied bath orbitals, the resulting all-electron wavefunction would be very close to the target solution and would converge quickly after another a few all-electron R-DIIS cycles as the second step. To obtain loosely converged DIIS orbitals, one could utilize techniques like smearing and level-shift \cite{Schlegel1991}, and loosen the criterion of convergence, e.g. by setting the energy difference between two sequential iterations as $\delta E<5\times 10^{-6}$ and the norm of orbital gradient $\Vert\mathbf{g}\Vert<0.01$, which was used in this study, in contrast to the default values of $\delta E< 10^{-9}$ and $\Vert\mathbf{g}\Vert<3.2\times 10^{-5}$, respectively.

\begin{figure}[!ht]
	\centering
	\includegraphics[width=\textwidth]{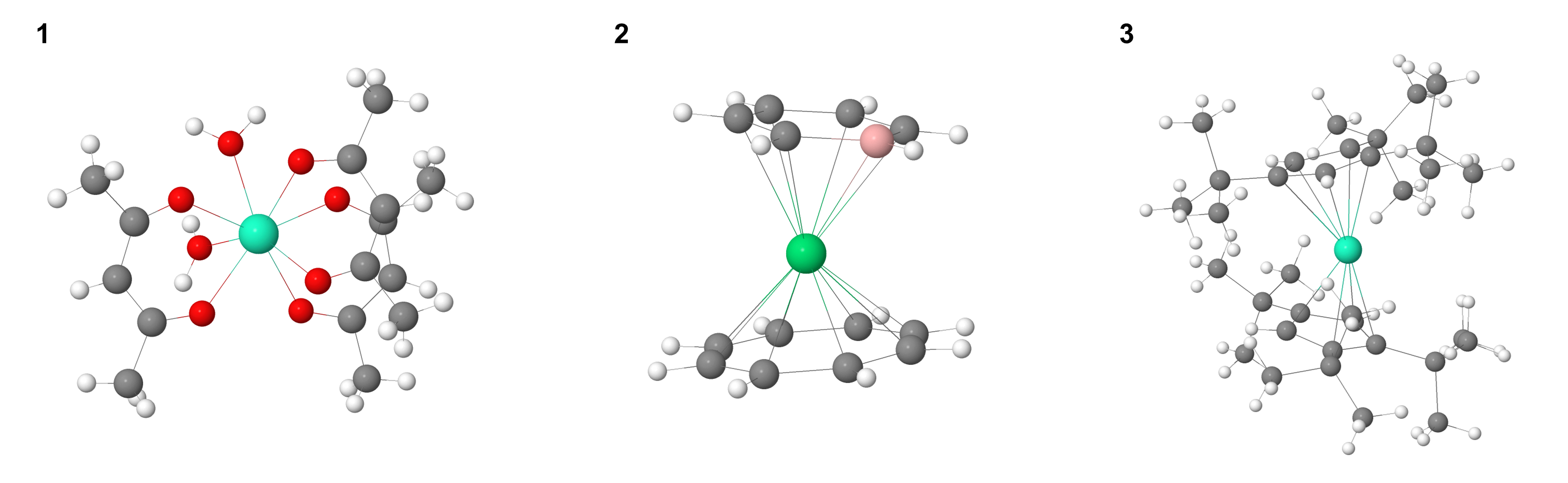}
 	\caption{Structures of the Lanthanide SIMs studied in this work extracted from Refs. \citenum{Briganti2021,Meng20162,Goodwin2017}. \textbf{1}: ${\rm Dy(acac)_3(H_{2}O)_2}$ (\textbf{1Dy} for short), mint/red/gray/white represent Dy/O/C/H atoms, respectively; \textbf{2}: ${\rm (BCp)Er(COT)}$ (\textbf{2Er} for short), green/gray/pink/white represent Er/C/B/H atoms, respectively; \textbf{3}: ${\rm [Dy(Cp^{ttt})_2]^+}$ (\textbf{3Dy} for short), mint/gray/white represent Dy/C/H atoms, respectively.}
	\label{fgr:4f-SIMs}
\end{figure}

\subsection{2.3. Computational Details}

We consider three prototypical 4f-SIMs, abbreviated as \textbf{1Dy}, \textbf{2Er} and \textbf{3Dy}, respectively, and illustrated in Figure~\ref{fgr:4f-SIMs}. Molecular structures extracted from experimentally measured crystal structures reported in the corresponding references \cite{Briganti2021, Meng20162, Goodwin2017} are used in our calculations. Scalar relativity and SOC effects are considered by using respectively the exact two-component (X2C) Foldy-Wouthuysen Hamiltonian \cite{Liu2009} and the spin-orbit mean-field (SOMF) approximation to the Breit-Pauli Hamiltonian \cite{Hess1996,Neese2005}. The Cholesky decomposition \cite{Aquilante2007} was used to reduce the memory requirement to store the molecular integrals. The basis sets used in the calculations are ANO-RCC-VTZP/VTZP/VDZP \cite{Widmark1990,Roos2004,Roos2008} for Dy/O/other atoms in \textbf{1Dy}, ANO-RCC-VTZP/VDZP for Er,B/other atoms in \textbf{2Er},  and ANO-RCC-VTZP/VDZP for Dy, the nearest C and other atoms in \textbf{3Dy}, respectively. For SA-CASSCF, we choose (9e,7o) and (11e,7o) as the CAS for Dy and Er complexes, respectively, and consider all spin multiplets including 490 doublets, 224 quartets and 21 sextets for \textbf{1Dy} and \textbf{3Dy}, and 112 doublets and 35 quartets for \textbf{2Er}. When building the DMET embedded Hamiltonian, we use $\varepsilon=10^{-5}$ to select bath orbitals. We have also conducted DMET+CASSI-SO calculation of \textbf{1Dy} using a more stringent criterion $\varepsilon = 10 ^{-13}$, and obtained essentially same results.

\begin{figure}[!ht]
    \centering
    \includegraphics[width=0.5\textwidth]{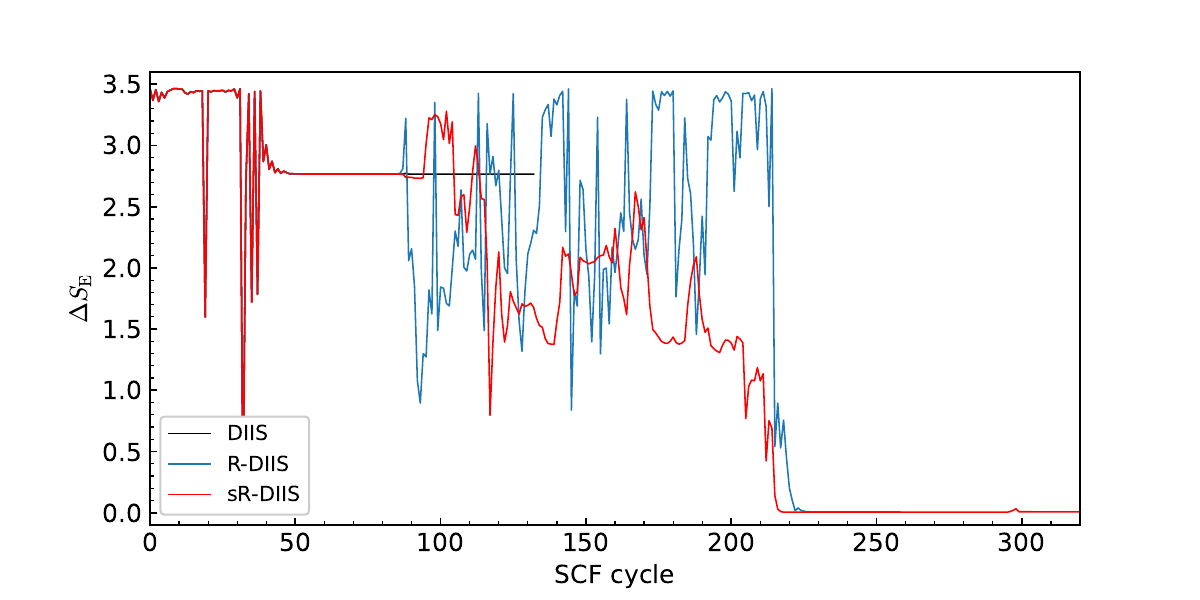}
    \includegraphics[width=0.5\textwidth]{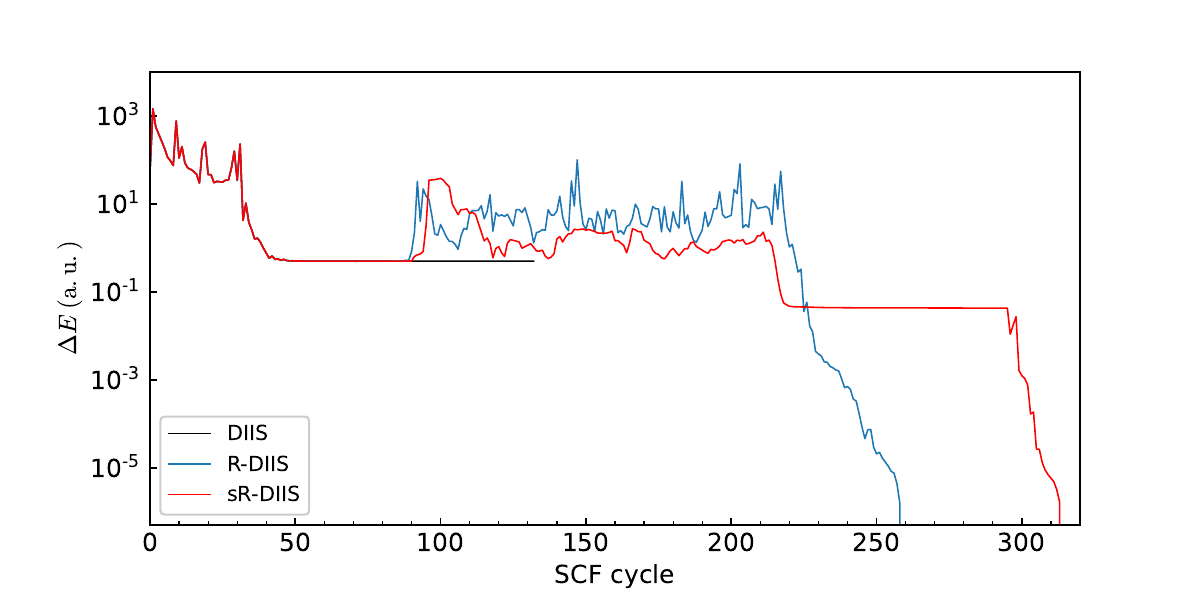}
    \caption{$\Delta S_\mathrm{E}$ and the deviation of the ROHF energy from the best-known minimum $\Delta E^*=E-E_\mathrm{conv}^*$ at each SCF cycle for \textbf{1Dy} using the default DIIS, R-DIIS and sR-DIIS. The latter two algorithms start on top of a loosely converged DIIS, and therefore the results of three schemes are the same for the first 80 cycles.}
    \label{fig:1Dy-DIIS}
\end{figure}

\section{3. Results and Discussion}

\subsection{3.1. R-DIIS and sR-DIIS for lanthanide complexes}
We take \textbf{1Dy} as an example to illustrate the performance of R-DIIS and sR-DIIS. For all ROHF SCF calculations, we employ the default ``minAO'' setting in PySCF as the initial guess unless otherwise specified. Figure~\ref{fig:1Dy-DIIS} shows $\Delta S_\mathrm{E}$ and the deviation of the ROHF energy from the best-known minimum, $\Delta E^* \equiv E - E_\mathrm{conv}^*$, plotted against the number of elapsed SCF cycles for DIIS, R-DIIS, and sR-DIIS. In the case of DIIS, $\Delta E^*$ changes rapidly during the first 50 cycles and then decreases smoothly until reaching a converged value of $\Delta E^* = 0.50$ a.u. after approximately 140 cycles. The final ROHF solution yields $\Delta S_\mathrm{E}^\mathrm{conv} = 2.766$, indicating that the singly occupied molecular orbitals (SOMOs) obtained from DIIS exhibit significant distribution over the ligand groups surrounding Dy$^{3+}$. For R-DIIS, the regularization term is activated after a loose DIIS convergence is achieved at around 80 cycles, effectively steering the SCF iteration away from the meta-stable DIIS solution. In the subsequent cycles, $\Delta S_\mathrm{E}$ remains significant, causing strong oscillations in the total energy. After approximately 120 additional cycles, $\Delta S_\mathrm{E}$ becomes vanishingly small, and the total energy rapidly decreases. Ultimately, a ROHF solution with a significantly lower energy than that from DIIS is obtained, with $\Delta S_\mathrm{E}^\mathrm{conv} = 0.007$. Similar trends are observed with the sR-DIIS technique, where the regularization term is applied to the error vector for the ROHF SCF iteration within the DMET subspace. Although sR-DIIS requires slightly more SCF cycles to reach convergence compared to R-DIIS, the computational cost is significantly reduced because the SCF iteration is performed in the much smaller DMET subspace. This reduction in dimensionality also allows all two-electron integrals involved in the subspace ROHF to be stored in memory rather than computed on the fly or loaded from disk, further enhancing the computational efficiency of sR-DIIS.

\begin{figure}[!ht]
	\centering
	\includegraphics[width=\textwidth]{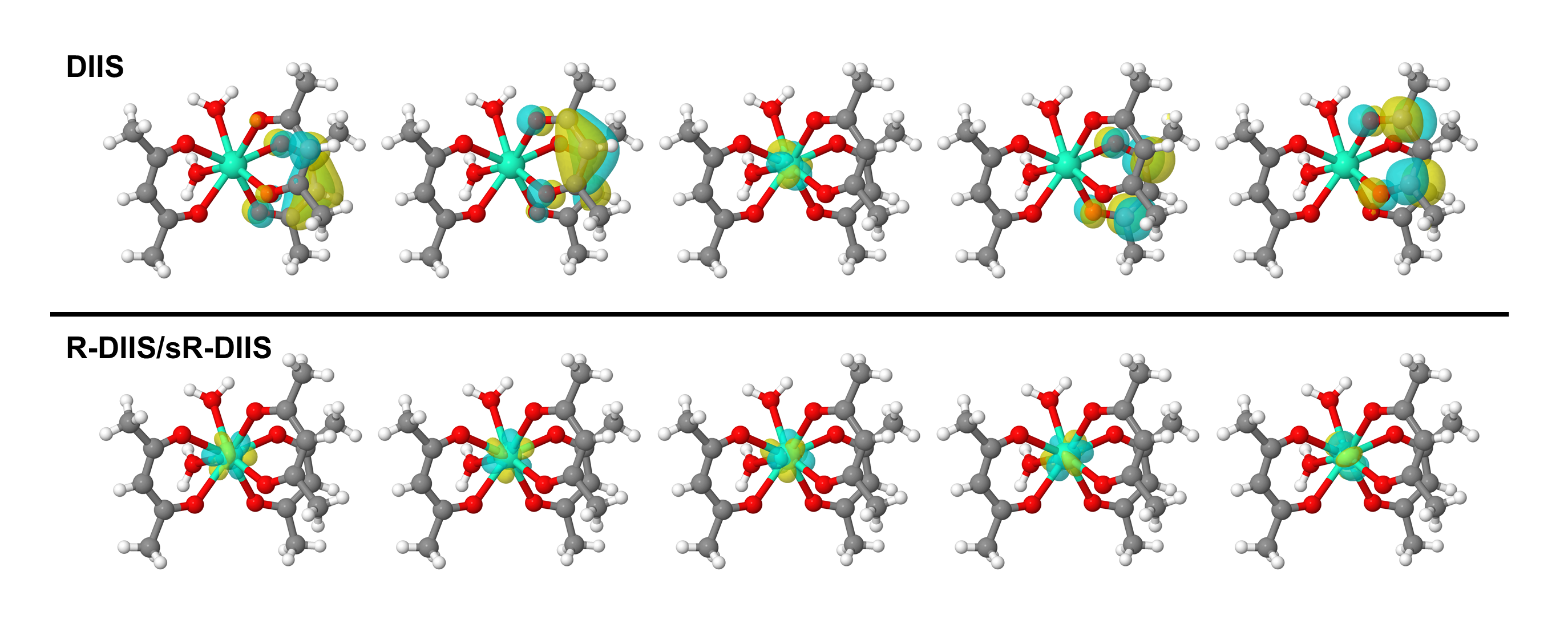}
	\caption{Contour plots of singly occupied molecular orbitals of \textbf{1Dy} molecule obtained from DIIS and R-DIIS/sR-DIIS ROHF.}
	\label{fig:1Dy-SOMOs}
  \end{figure}

To more clearly demonstrate the effects of R-DIIS on ROHF solutions for lanthanide complexes, we visualize five SOMOs of \textbf{1Dy} in the DIIS and R-DIIS solutions in Figure~\ref{fig:1Dy-SOMOs}. In the DIIS solution, only one SOMO is strongly localized on the Dy atom with apparent 4f character, and other four SOMOs are mainly distributed in the surrounding ligand groups. In contrast, the R-DIIS solution has all unpaired electrons strongly localized in the Dy 4f orbitals. This suggests that the wavefunctions obtained through R-DIIS are more suitable as starting points for DMET. Additionally, the fact that the R-DIIS solution has a lower energy than the DIIS solution indicates that it is closer to the true global minimum ROHF solution. Consequently, R-DIIS wavefunctions may outperform DIIS solutions not only in absolute or relative energy calculations outside of DMET but also in providing more accurate mean-field-level chemical insights, such as population analysis. Furthermore, they could serve as better initial points for all-electron correlated methods, such as coupled cluster theory. These potential advantages will be explored in future research.

\begin{table}[!ht]
	\caption{Energies (in $\rm cm^{-1}$) of 11 lowest pre-SOC states corresponding to $^6\mathrm{H}$  of \textbf{1Dy} obtained from DMET-based SA-CASSCF using DIIS and R-DIIS converged ROHF solution as the starting point compared to those from all-electron (AE) SA-CASSCF.}
	\begin{tabular}{ccccc}
		\hline
	             & \multicolumn{2}{c}{DIIS-based} &\multicolumn{2}{c}{R-DIIS-based} \\
		State    &DMET &  AE &  DMET & AE \\
		\hline
		1 &   0.0   & 0.0    & 0.0   &   0.0 \\
		2 & 410.9   & 47.7   &   5.9 &   6.1 \\
		3 & 3532.3  & 4967.1 & 208.7 & 202.4 \\
		4 & 3993.2  & 5181.0 & 272.5 & 270.4 \\
		5 & 5398.4  & 6439.5 & 300.4 & 297.5 \\
		6 & 6109.0  & 6499.7 & 376.8 & 367.2 \\
		7 & 8889.3  & 11208.6& 409.4 & 401.2 \\
		8 & 9581.7  & 11398.4& 526.3 & 513.1 \\
		9 & 42632.2 & 45795.0& 537.5 & 524.2 \\
		10& 43068.3 & 45806.0& 617.7 & 602.5 \\
		11& 48000.3 & 50628.5& 623.8 & 608.7 \\
		\hline
	\end{tabular}
	\label{tab:1Dy-pre-SOC}
\end{table}

\begin{figure}[!ht]
	\centering
	\includegraphics[width=0.5\textwidth]{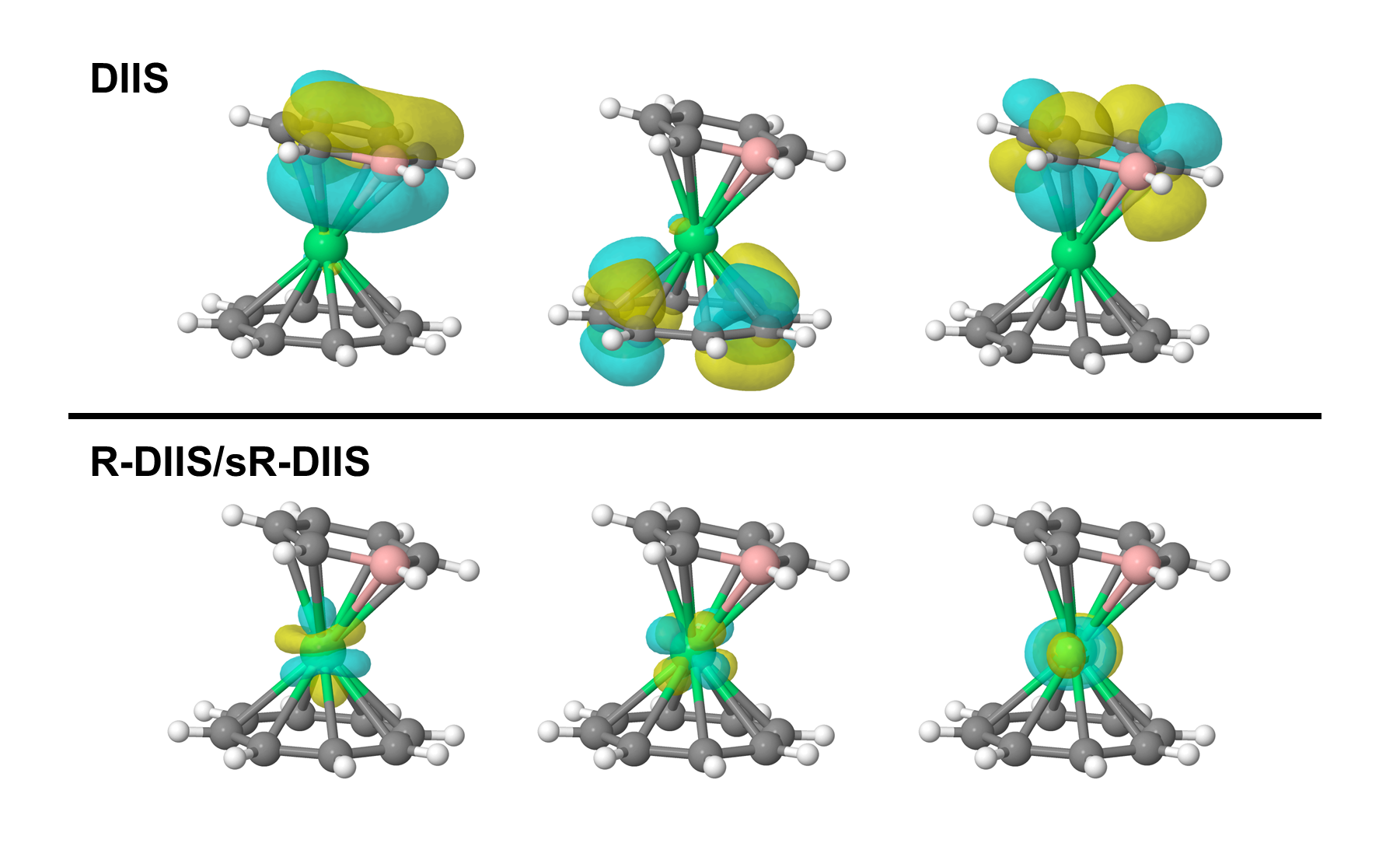}
	\caption{Contour plots of singly occupied molecular orbitals of \textbf{2Er} obtained from DIIS and R-DIIS/sR-DIIS ROHF.}
	\label{fig:2Er-SOMOs}
  \end{figure}

\begin{table}[!ht]
	\caption{Energies (in $\rm cm^{-1}$) of 13 lowest pre-SOC states corresponding to $^4\mathrm{I}$  of \textbf{2Er} obtained from DMET-based SA-CASSCF using DIIS and R-DIIS converged ROHF solution as the starting point compared to those from all-electron (AE) SA-CASSCF.}
	\begin{tabular}{ccccc}
		\hline
		         & \multicolumn{2}{c}{DIIS-based} &\multicolumn{2}{c}{R-DIIS-based} \\
		State    &DMET   & AE    & DMET & AE  \\
		\hline
		1 & 0.0    & 0.0   & 0.0    & 0.0 \\
		2 & 0.2    & 0.1   & 0.1    & 0.1 \\
		3 & 127.9  & 159.5 & 159.7  & 159.5 \\
		4 & 143.6  & 162.0 & 161.8  & 162.0 \\
		5 & 189.8  & 176.0 & 176.3  & 176.0 \\
		6 & 191.7  & 186.4 & 186.1  & 186.4 \\
		7 & 233.7  & 225.8 & 226.3  & 225.8 \\
		8 & 280.7  & 234.6 & 234.7  & 234.6 \\
		9 & 292.9  & 238.4 & 238.7  & 238.4 \\
		10 & 333.5 & 267.9 & 268.7  & 267.9 \\
		11 & 343.8 & 276.2 & 277.1  & 276.2 \\
		12 & 435.3 & 308.3 & 309.0  & 308.3 \\
		13 & 436.9 & 309.9 & 310.5  & 310.0 \\
		\hline
	\end{tabular}
	\label{tab:2Er-pre-SOC}
\end{table}

\subsection{3.2. Importance of obtaining physically correct ROHF solution}

In our previous work on 3d-SIMs \cite{Ai2022}, we found that the efficacy of DMET strongly depends on the quality of the ROHF solution that is used to build the  embedded cluster space, and using the meta-stable ROHF solution obtained from the default DIIS technique leads to significant discrepancies between DMET and all-electron (AE) results. To clearly reveal the importance of using physically correct low-level wavefunctions for DMET treatment of lanthanide systems, we compare in Table~\ref{tab:1Dy-pre-SOC} and Table~\ref{tab:2Er-pre-SOC} the energies of 11 and 13 lowest pre-SOC states of \textbf{1Dy} and \textbf{2Er}, corresponding to the ground state spin multiplet $^6\mathrm{H}$ and $^4\mathrm{I}$, respectively, obtained from DMET-based SA-CASSCF on top of physically correct (R-DIIS) or meta-stable (DIIS) ROHF solutions to those obtained from all-electron treatments. We can see that \textbf{1Dy} and \textbf{2Er} show different features, representing two typical scenarios in lanthanide complexes. For \textbf{1Dy}, the results of all-electron SA-CASSCF strongly depend on whether using DIIS or R-DIIS ROHF solutions as the starting point. The former leads to the energies of the crystal-field split states of $^6\mathrm{H}$ falling an unphysical range of $10^4$ cm$^{-1}$, indicating an ill-behaved CASSCF calculation as a result of poor initial guess. In contrast, the corresponding results obtained from using the R-DIIS solution as the starting point fall in a range of a few hundred cm$^{-1}$, which is typical for lanthanide complexes \cite{Ungur2015}. The DMET results for \textbf{1Dy} also show dramatic discrepancies when using DIIS or R-DIIS solution as the starting point, and the R-DIIS based DMET leads to good agreement with the all-electron results with a relative error of less than 3\%. In contrast, DIIS-based DMET results show significant differences from all-electron ones and still fall in a unphysical range.

The results for \textbf{2Er} show some different features. While using the default DIIS still leads to a metastable ROHF solution with dominantly delocalized spin polarization on ligand groups, as shown in Figure \ref{fig:2Er-SOMOs}, all-electron SA-CASSCF with DIIS and R-DIIS solutions as the starting point give nearly identical results. But there are significant discrepancies in the DMET results based on DIIS and R-DIIS ROHF solution, and the maximum difference between the energies of crystal-field splitting states of $^4$I reaches about 120 cm$^{-1}$. Using the R-DIIS solution to build the embedded cluster space, the differences between DMET and all-electron CASSCF results are less than 1 cm$^{-1}$. It is therefore clear that using meta-stable low-level wavefunctions obtained from the original DIIS may yield inappropriate bath orbitals, or equivalently speaking, inappropriate unentangled occupied (core) orbitals that are frozen during the orbital rotation procedure when SA-CASSCF is applied to the embedded cluster space, leading to results that differ significantly from all-electron results. Therefore, we demonstrate that physically correct low-level wavefunctions acquired by R-DIIS method are indeed crucial for the efficacy of DMET.

We should also mention that using R-DIIS ROHF solution as the initial guess for SA-CASSCF can also significantly reduce the number of iterations required to attain self-consistency. In our calculations, we found that for \textbf{1Dy}, SA-CASSCF starting from the DIIS solution takes 40 iterations (rotation of orbitals) to converge, and that from the R-DIIS solution takes only 4 iterations. For \textbf{2Er}, the number of SA-CASSCF iterations are 12 and 4 for DIIS and R-DIIS based all-electron SA-CASSCF, respectively. These results clearly show the importance of using the physically correction ROHF solution, ensured by using R-DIIS technique, not only for building an accurate embedded cluster Hamiltonian that is crucial for the overall performance of the DMET approach, but also for providing more robust initial guess for CASSCF calculation. Therefore we expect that the R-DIIS/sR-DIIS technique will play important roles in a wide range of applications that involve transition metal, lanthanide or actinide metal centers with unpaired $d$ or $f$ electrons.

\begin{table}[!ht]
	\caption{Lowest SOC state energies ($\rm cm^{-1}$) of \textbf{1Dy}, \textbf{2Er} and \textbf{3Dy} obtained from DMET and all-electron (AE) CASSI-SO calculation. Each state here refers to a degenerate pair of Kramers doublet. The second row shows the number of orbitals used in DMET or AE Hamiltonian. The last row shows the root mean square error (RMSE) of DMET results with respect to AE ones.
	}
	\begin{tabular}{crrrrrr}
		\hline\hline
		& \multicolumn{2}{c}{\textbf{1Dy}} & \multicolumn{2}{c}{\textbf{2Er}} & \multicolumn{2}{c}{\textbf{3Dy}} \\
		State/$N_{\rm orb}$
		  & DMET  & AE    & DMET  & AE     & DMET   & AE \\ \hline
		$N_{\rm orb}$
		  & 170   & 679   & 164   & 386    & 174    & 1030  \\ \hline
		1 & 0.0   & 0.0   & 0.0   & 0.0    & 0.0    & 0.0    \\
		2 & 157.4 & 154.8 & 147.8 & 147.5  & 512.9  & 508.7  \\
		3 & 237.0 & 233.0 & 199.4 & 199.1  & 818.4  & 810.2  \\
		4 & 292.0 & 287.4 & 211.6 & 211.1  & 1021.9 & 1011.1 \\
		5 & 332.6 & 322.3 & 239.6 & 238.6  & 1200.8 & 1188.3 \\
		6 & 426.9 & 413.9 & 249.3 & 248.7  & 1365.7 & 1352.1 \\
		7 & 484.0 & 471.0 & 269.3 & 268.4  & 1493.3 & 1479.5 \\
		8 & 553.1 & 537.0 & 295.7 & 294.7  & 1575.8 & 1559.3 \\
	\hline
		RMSE & 9.7& 0     & 0.7   & 0      & 11.2   & 0    \\ \hline\hline
	\end{tabular}
	\label{tab:4f-SIM}
\end{table}

\subsection{3.3. Performances of DMET+CASSI-SO for 4f-SIMs}
As a direct demonstration of the utility of the DMET method for lanthanide SIMs, we present the energies of the lowest Kramers doublets corresponding to crystal field splitting of the ground state multiplet calculated by all-electron and DMET based CASSI-SO for \textbf{1Dy}, \textbf{2Er} and \textbf{3Dy} in Table~\ref{tab:4f-SIM}. For all these three lanthanide complexes, we see very good agreement between DMET and all-electron results. For \textbf{2Er}, the root mean square error (RMSE) of DMET results with respect to all-electron ones is only 0.7 cm$^{-1}$. The errors of DMET with respect to the all-electron treatment are lightly larger in \textbf{1Dy} and \textbf{3Dy}, with RMSE of about 10 cm$^{-1}$, which, however, accounts for a negligibly small relative error of only about 3\%.
Considering the dramatic reduction of the number of orbitals involved in the embedded cluster Hamiltonian with respect to the full Hamiltonian, the accuracy that can be achieved by DMET is quite remarkable. It is also noteworthy that the size of the embedded cluster space is essentially independent of the size of ligand groups, which makes DMET+CASSI-SO particularly promising for theoretical study of lanthanide SIMs with more complex bulky ligand groups that have been synthesized experimentally (see, e.g. Refs.\citenum{Wang2024, Wang2023, Feng2018}) or SIMs deposited on surfaces that are key to practical applications \cite{Cinchetti2017}.


\section{Conclusions}

In conclusion, in this work we have extended DMET+CASSI-SO that is previously developed for 3d SIMs to lanthanide SIM systems. We have presented a detailed formulation of the regularized direct inversion of iterative subspace (R-DIIS) algorithm, which guarantees obtaining physically correct restricted open-shell Hartree-Fock (ROHF) wavefunctions that is crucial to the efficacy of DMET. Furthermore, we have introduced the subspace R-DIIS (sR-DIIS) algorithm, which demonstrates enhanced efficiency and reliability for lanthanide systems. Through extensive testing on representative lanthanide single-ion magnets (4f-SIMs), we have validated the performance of these algorithms and showcased the remarkable accuracy of the DMET+CASSI-SO approach. We believe that this improved methodology will open new avenues for large-scale theoretical investigations of complex lanthanide systems, offering both computational efficiency and high precision. We also envisage that SCF wavefunctions obtained by R-DIIS/sR-DIIS might be useful outside the regime of quantum embedding, for example, in the DFT calculation of high-spin/low-spin energy differences or as an starting point of the all-electron correlated methods like selected configuration interaction \cite{Tubman2016,Sharma2017,Liu2016} or density matrix renormalization group \cite{White1992,ChanGKL2011} that requires an active space.

\begin{acknowledgement}
This work is partly supported by the National Natural Science Foundation of China (Project Number 12234001). We acknowledge the High-performance Computing Platform of Peking University for providing the computational facility.
\end{acknowledgement}


\begin{mcitethebibliography}{70}
\providecommand*\natexlab[1]{#1}
\providecommand*\mciteSetBstSublistMode[1]{}
\providecommand*\mciteSetBstMaxWidthForm[2]{}
\providecommand*\mciteBstWouldAddEndPuncttrue
  {\def\EndOfBibitem{\unskip.}}
\providecommand*\mciteBstWouldAddEndPunctfalse
  {\let\EndOfBibitem\relax}
\providecommand*\mciteSetBstMidEndSepPunct[3]{}
\providecommand*\mciteSetBstSublistLabelBeginEnd[3]{}
\providecommand*\EndOfBibitem{}
\mciteSetBstSublistMode{f}
\mciteSetBstMaxWidthForm{subitem}{(\alph{mcitesubitemcount})}
\mciteSetBstSublistLabelBeginEnd
  {\mcitemaxwidthsubitemform\space}
  {\relax}
  {\relax}

\bibitem[Seijo \latin{et~al.}(2016)Seijo, Barandiarán, Bünzli, and
  Pecharsky]{Seijo2016}
Seijo,~L.; Barandiarán,~Z.; Bünzli,~J.-C.~G.; Pecharsky,~V.~K. \emph{Handbook
  on the Physics and Chemistry of Rare Earths}; Elsevier, 2016; Vol.~50; pp
  65--89\relax
\mciteBstWouldAddEndPuncttrue
\mciteSetBstMidEndSepPunct{\mcitedefaultmidpunct}
{\mcitedefaultendpunct}{\mcitedefaultseppunct}\relax
\EndOfBibitem
\bibitem[Dolg(2015)]{Dolg2015}
Dolg,~M. \emph{Computational Methods in Lanthanide and Actinide Chemistry};
  John Wiley \& Sons, Ltd, 2015\relax
\mciteBstWouldAddEndPuncttrue
\mciteSetBstMidEndSepPunct{\mcitedefaultmidpunct}
{\mcitedefaultendpunct}{\mcitedefaultseppunct}\relax
\EndOfBibitem
\bibitem[Reid \latin{et~al.}(2016)Reid, Bünzli, and Pecharsky]{Reid2016}
Reid,~M.~F.; Bünzli,~J.-C.~G.; Pecharsky,~V.~K. \emph{Handbook on the Physics
  and Chemistry of Rare Earths}; Elsevier, 2016; Vol.~50; pp 47--64\relax
\mciteBstWouldAddEndPuncttrue
\mciteSetBstMidEndSepPunct{\mcitedefaultmidpunct}
{\mcitedefaultendpunct}{\mcitedefaultseppunct}\relax
\EndOfBibitem
\bibitem[Barandir{'a}n \latin{et~al.}(2022)Barandir{'a}n, Joos, and
  Seijo]{Barandiaran2022}
Barandir{'a}n,~Z.; Joos,~J.; Seijo,~L. \emph{Luminescent Materials: A Quantum
  Chemical Approach for Computer-Aided Discovery and Design}; Springer,
  2022\relax
\mciteBstWouldAddEndPuncttrue
\mciteSetBstMidEndSepPunct{\mcitedefaultmidpunct}
{\mcitedefaultendpunct}{\mcitedefaultseppunct}\relax
\EndOfBibitem
\bibitem[Ogasawara and Watanabe(2008)Ogasawara, and Watanabe]{Ogasawara2008}
Ogasawara,~K.; Watanabe,~S. \emph{Advances in Quantum Chemistry}; Academic
  Press, 2008; Vol.~54; pp 297--314\relax
\mciteBstWouldAddEndPuncttrue
\mciteSetBstMidEndSepPunct{\mcitedefaultmidpunct}
{\mcitedefaultendpunct}{\mcitedefaultseppunct}\relax
\EndOfBibitem
\bibitem[Jia \latin{et~al.}(2016)Jia, Miglio, Ponc{\'{e}}, Gonze, and
  Mikami]{Jia2016}
Jia,~Y.; Miglio,~A.; Ponc{\'{e}},~S.; Gonze,~X.; Mikami,~M. First-principles
  study of \ce{Ce^3+}-doped lanthanum silicate nitride phosphors: Neutral
  excitation, Stokes shift, and luminescent center identification. \emph{Phys.
  Rev. B} \textbf{2016}, \emph{93}, 155111\relax
\mciteBstWouldAddEndPuncttrue
\mciteSetBstMidEndSepPunct{\mcitedefaultmidpunct}
{\mcitedefaultendpunct}{\mcitedefaultseppunct}\relax
\EndOfBibitem
\bibitem[Filatov and Shaik(1998)Filatov, and Shaik]{Filatov1998}
Filatov,~M.; Shaik,~S. Spin-restricted density functional approach to the
  open-shell problem. \emph{Chem. Phys. Lett.} \textbf{1998}, \emph{288},
  689--697\relax
\mciteBstWouldAddEndPuncttrue
\mciteSetBstMidEndSepPunct{\mcitedefaultmidpunct}
{\mcitedefaultendpunct}{\mcitedefaultseppunct}\relax
\EndOfBibitem
\bibitem[Frank \latin{et~al.}(1998)Frank, Hutter, Marx, and
  Parrinello]{Frank1998}
Frank,~I.; Hutter,~J.; Marx,~D.; Parrinello,~M. Molecular dynamics in low-spin
  excited states. \emph{J. Chem. Phys.} \textbf{1998}, \emph{108},
  689--697\relax
\mciteBstWouldAddEndPuncttrue
\mciteSetBstMidEndSepPunct{\mcitedefaultmidpunct}
{\mcitedefaultendpunct}{\mcitedefaultseppunct}\relax
\EndOfBibitem
\bibitem[Ramanantoanina \latin{et~al.}(2015)Ramanantoanina, Sahnoun, Barbiero,
  Ferbinteanu, and Cimpoesu]{Ramanantoanina2015}
Ramanantoanina,~H.; Sahnoun,~M.; Barbiero,~A.; Ferbinteanu,~M.; Cimpoesu,~F.
  Development and applications of the LFDFT: the non-empirical account of
  ligand field and the simulation of the f-d transitions by density functional
  theory. \emph{Phys. Chem. Chem. Phys.} \textbf{2015}, \emph{17},
  18547--18557\relax
\mciteBstWouldAddEndPuncttrue
\mciteSetBstMidEndSepPunct{\mcitedefaultmidpunct}
{\mcitedefaultendpunct}{\mcitedefaultseppunct}\relax
\EndOfBibitem
\bibitem[Roos \latin{et~al.}(2016)Roos, Lindh, Malmqvist, Veryazov, and
  Widmark]{Roos2016}
Roos,~B.~O.; Lindh,~R.; Malmqvist,~P.~{\AA}.; Veryazov,~V.; Widmark,~P.-O.
  \emph{Multiconfigurational Quantum Chemistry}; John Wiley \& Sons, Ltd,
  2016\relax
\mciteBstWouldAddEndPuncttrue
\mciteSetBstMidEndSepPunct{\mcitedefaultmidpunct}
{\mcitedefaultendpunct}{\mcitedefaultseppunct}\relax
\EndOfBibitem
\bibitem[Roos \latin{et~al.}(1980)Roos, Taylor, and Siegbahn]{Roos1980}
Roos,~B.~O.; Taylor,~P.~R.; Siegbahn,~P. E.~M. A Complete Active Space SCF
  Method (CASSCF) Using a Density-matrix Formulated Super-CI Approach.
  \emph{Chem. Phys.} \textbf{1980}, \emph{48}, 157--173\relax
\mciteBstWouldAddEndPuncttrue
\mciteSetBstMidEndSepPunct{\mcitedefaultmidpunct}
{\mcitedefaultendpunct}{\mcitedefaultseppunct}\relax
\EndOfBibitem
\bibitem[Olsen \latin{et~al.}(1988)Olsen, Roos, J{\o}rgensen, and
  Jensen]{Olsen1988}
Olsen,~J.; Roos,~B.~O.; J{\o}rgensen,~P.; Jensen,~H. J.~A. Determinant Based
  Configuration-interaction Algorithms for Complete and Restricted
  Configuration-interaction Spaces. \emph{J. Chem. Phys.} \textbf{1988},
  \emph{89}, 2185--2192\relax
\mciteBstWouldAddEndPuncttrue
\mciteSetBstMidEndSepPunct{\mcitedefaultmidpunct}
{\mcitedefaultendpunct}{\mcitedefaultseppunct}\relax
\EndOfBibitem
\bibitem[Ma \latin{et~al.}(2011)Ma, Li, and Gagliardi]{MaD2011}
Ma,~D.; Li,~M.~G.; Gagliardi,~L. The Generalized Active Space Concept in
  Multiconfigurational Self-consistent Field Methods. \emph{J. Chem. Phys.}
  \textbf{2011}, \emph{135}, 044128\relax
\mciteBstWouldAddEndPuncttrue
\mciteSetBstMidEndSepPunct{\mcitedefaultmidpunct}
{\mcitedefaultendpunct}{\mcitedefaultseppunct}\relax
\EndOfBibitem
\bibitem[Lischka \latin{et~al.}(2018)Lischka, Nachtigallova, Aquino, Szalay,
  Plasser, Machado, and Barbatti]{Lischka2018}
Lischka,~H.; Nachtigallova,~D.; Aquino,~A. J.~A.; Szalay,~P.~G.; Plasser,~F.;
  Machado,~F. B.~C.; Barbatti,~M. Multireference {Approaches} for {Excited}
  {States} of {Molecules}. \emph{Chem. Rev.} \textbf{2018}, \emph{118},
  7293--7361\relax
\mciteBstWouldAddEndPuncttrue
\mciteSetBstMidEndSepPunct{\mcitedefaultmidpunct}
{\mcitedefaultendpunct}{\mcitedefaultseppunct}\relax
\EndOfBibitem
\bibitem[Park \latin{et~al.}(2020)Park, Al-Saadon, MacLeod, Shiozaki, and
  Vlaisavljevich]{Park2020}
Park,~J.~W.; Al-Saadon,~R.; MacLeod,~M.~K.; Shiozaki,~T.; Vlaisavljevich,~B.
  Multireference {Electron} {Correlation} {Methods}: {Journeys} along
  {Potential} {Energy} {Surfaces}. \emph{Chem. Rev.} \textbf{2020}, \emph{120},
  5878--5909\relax
\mciteBstWouldAddEndPuncttrue
\mciteSetBstMidEndSepPunct{\mcitedefaultmidpunct}
{\mcitedefaultendpunct}{\mcitedefaultseppunct}\relax
\EndOfBibitem
\bibitem[Atanasov \latin{et~al.}(2015)Atanasov, Aravena, Suturina, Bill,
  Maganas, and Neese]{Atanasov2015}
Atanasov,~M.; Aravena,~D.; Suturina,~E.; Bill,~E.; Maganas,~D.; Neese,~F. First
  Principles Approach to the Electronic Structure, Magnetic Anisotropy and Spin
  Relaxation in Mononuclear 3d-transition Metal Single Molecule Magnets.
  \emph{Coord. Chem. Rev.} \textbf{2015}, \emph{289}, 177--214\relax
\mciteBstWouldAddEndPuncttrue
\mciteSetBstMidEndSepPunct{\mcitedefaultmidpunct}
{\mcitedefaultendpunct}{\mcitedefaultseppunct}\relax
\EndOfBibitem
\bibitem[Woodruff \latin{et~al.}(2013)Woodruff, Winpenny, and
  Layfield]{Woodruff2013}
Woodruff,~D.~N.; Winpenny,~R. E.~P.; Layfield,~R.~A. Lanthanide
  {Single}-{Molecule} {Magnets}. \emph{Chem. Rev.} \textbf{2013}, \emph{113},
  5110--5148\relax
\mciteBstWouldAddEndPuncttrue
\mciteSetBstMidEndSepPunct{\mcitedefaultmidpunct}
{\mcitedefaultendpunct}{\mcitedefaultseppunct}\relax
\EndOfBibitem
\bibitem[Meng \latin{et~al.}(2016)Meng, Jiang, Wang, and Gao]{Meng2016}
Meng,~Y.~S.; Jiang,~S.~D.; Wang,~B.~W.; Gao,~S. Understanding the {Magnetic}
  {Anisotropy} toward {Single}-{Ion} {Magnets}. \emph{Acc. Chem. Res.}
  \textbf{2016}, \emph{49}, 2381--2389\relax
\mciteBstWouldAddEndPuncttrue
\mciteSetBstMidEndSepPunct{\mcitedefaultmidpunct}
{\mcitedefaultendpunct}{\mcitedefaultseppunct}\relax
\EndOfBibitem
\bibitem[Lucas \latin{et~al.}(2015)Lucas, Lucas, Mercier, Rollat, and
  Davenport]{Lucas2015}
Lucas,~J.; Lucas,~P.; Mercier,~T.~L.; Rollat,~A.; Davenport,~W.
  \emph{Introduction to Rare Earth Luminescent Materials}; Elsevier, 2015;
  Chapter 15, pp 251 -- 280\relax
\mciteBstWouldAddEndPuncttrue
\mciteSetBstMidEndSepPunct{\mcitedefaultmidpunct}
{\mcitedefaultendpunct}{\mcitedefaultseppunct}\relax
\EndOfBibitem
\bibitem[Sun and Jiang(2023)Sun, and Jiang]{Sun2023}
Sun,~H.-Y.; Jiang,~H. Combined {DFT} and wave function theory approach to
  excited states of lanthanide luminescent materials: A case study of
  LaF3:Ce3+. \emph{J. Ch. Chem. Soc.} \textbf{2023}, \emph{70}, 604--617\relax
\mciteBstWouldAddEndPuncttrue
\mciteSetBstMidEndSepPunct{\mcitedefaultmidpunct}
{\mcitedefaultendpunct}{\mcitedefaultseppunct}\relax
\EndOfBibitem
\bibitem[Sun and Chan(2016)Sun, and Chan]{Sun2016}
Sun,~Q.; Chan,~G. K.-L. Quantum {Embedding} {Theories}. \emph{Acc. Chem. Res.}
  \textbf{2016}, \emph{49}, 2705--2712\relax
\mciteBstWouldAddEndPuncttrue
\mciteSetBstMidEndSepPunct{\mcitedefaultmidpunct}
{\mcitedefaultendpunct}{\mcitedefaultseppunct}\relax
\EndOfBibitem
\bibitem[Knizia and Chan(2012)Knizia, and Chan]{Knizia2012}
Knizia,~G.; Chan,~G. K.-L. Density {Matrix} {Embedding}: {A} {Simple}
  {Alternative} to {Dynamical} {Mean}-{Field} {Theory}. \emph{Phys. Rev. Lett.}
  \textbf{2012}, \emph{109}, 186404\relax
\mciteBstWouldAddEndPuncttrue
\mciteSetBstMidEndSepPunct{\mcitedefaultmidpunct}
{\mcitedefaultendpunct}{\mcitedefaultseppunct}\relax
\EndOfBibitem
\bibitem[Wouters \latin{et~al.}(2016)Wouters, Jimenez-Hoyos, Sun, and
  Chan]{Wouters2016}
Wouters,~S.; Jimenez-Hoyos,~C.~A.; Sun,~Q.; Chan,~G. K.-L. A {Practical}
  {Guide} to {Density} {Matrix} {Embedding} {Theory} in {Quantum} {Chemistry}.
  \emph{J. Chem. Theory Comput.} \textbf{2016}, \emph{12}, 2706--2719\relax
\mciteBstWouldAddEndPuncttrue
\mciteSetBstMidEndSepPunct{\mcitedefaultmidpunct}
{\mcitedefaultendpunct}{\mcitedefaultseppunct}\relax
\EndOfBibitem
\bibitem[Pandharkar \latin{et~al.}(2019)Pandharkar, Hermes, Cramer, and
  Gagliardi]{Pandharkar2019}
Pandharkar,~R.; Hermes,~M.~R.; Cramer,~C.~J.; Gagliardi,~L. Spin-{State}
  {Ordering} in {Metal}-{Based} {Compounds} {Using} the {Localized} {Active}
  {Space} {Self}-{Consistent} {Field} {Method}. \emph{J. Phys. Chem. Lett.}
  \textbf{2019}, \emph{10}, 5507--5513\relax
\mciteBstWouldAddEndPuncttrue
\mciteSetBstMidEndSepPunct{\mcitedefaultmidpunct}
{\mcitedefaultendpunct}{\mcitedefaultseppunct}\relax
\EndOfBibitem
\bibitem[Ai \latin{et~al.}(2022)Ai, Sun, and Jiang]{Ai2022}
Ai,~Y.; Sun,~Q.; Jiang,~H. Efficient Multi-configurational Quantum Chemistry
  Approach to Single-ion Magnets Based on Density Matrix Embedding Theory.
  \emph{J. Phys. Chem. Lett.} \textbf{2022}, \emph{13}, 10627--10634\relax
\mciteBstWouldAddEndPuncttrue
\mciteSetBstMidEndSepPunct{\mcitedefaultmidpunct}
{\mcitedefaultendpunct}{\mcitedefaultseppunct}\relax
\EndOfBibitem
\bibitem[Pham \latin{et~al.}(2018)Pham, Bernales, and Gagliardi]{Pham2018}
Pham,~H.~Q.; Bernales,~V.; Gagliardi,~L. Can {Density} {Matrix} {Embedding}
  {Theory} with the {Complete} {Activate} {Space} {Self}-{Consistent} {Field}
  {Solver} {Describe} {Single} and {Double} {Bond} {Breaking} in {Molecular}
  {Systems}? \emph{J. Chem. Theory Comput.} \textbf{2018}, \emph{14},
  1960--1968\relax
\mciteBstWouldAddEndPuncttrue
\mciteSetBstMidEndSepPunct{\mcitedefaultmidpunct}
{\mcitedefaultendpunct}{\mcitedefaultseppunct}\relax
\EndOfBibitem
\bibitem[Hermes and Gagliardi(2019)Hermes, and Gagliardi]{Hermes2019}
Hermes,~M.~R.; Gagliardi,~L. Multiconfigurational {Self}-{Consistent} {Field}
  {Theory} with {Density} {Matrix} {Embedding}: {The} {Localized} {Active}
  {Space} {Self}-{Consistent} {Field} {Method}. \emph{J. Chem. Theory Comput.}
  \textbf{2019}, \emph{15}, 972--986\relax
\mciteBstWouldAddEndPuncttrue
\mciteSetBstMidEndSepPunct{\mcitedefaultmidpunct}
{\mcitedefaultendpunct}{\mcitedefaultseppunct}\relax
\EndOfBibitem
\bibitem[Hermes \latin{et~al.}(2020)Hermes, Pandharkar, and
  Gagliardi]{Hermes2020}
Hermes,~M.~R.; Pandharkar,~R.; Gagliardi,~L. Variational {Localized} {Active}
  {Space} {Self}-{Consistent} {Field} {Method}. \emph{J. Chem. Theory Comput.}
  \textbf{2020}, \emph{16}, 4923--4937\relax
\mciteBstWouldAddEndPuncttrue
\mciteSetBstMidEndSepPunct{\mcitedefaultmidpunct}
{\mcitedefaultendpunct}{\mcitedefaultseppunct}\relax
\EndOfBibitem
\bibitem[Mitra \latin{et~al.}(2021)Mitra, Pham, Pandharkar, Hermes, and
  Gagliardi]{Mitra2021}
Mitra,~A.; Pham,~H.~Q.; Pandharkar,~R.; Hermes,~M.~R.; Gagliardi,~L. Excited
  {States} of {Crystalline} {Point} {Defects} with {Multireference} {Density}
  {Matrix} {Embedding} {Theory}. \emph{J. Phys. Chem. Lett.} \textbf{2021},
  \emph{12}, 11688--11694\relax
\mciteBstWouldAddEndPuncttrue
\mciteSetBstMidEndSepPunct{\mcitedefaultmidpunct}
{\mcitedefaultendpunct}{\mcitedefaultseppunct}\relax
\EndOfBibitem
\bibitem[Mitra \latin{et~al.}(2022)Mitra, Hermes, Cho, Agarawal, and
  Gagliardi]{Mitra2022}
Mitra,~A.; Hermes,~M.~R.; Cho,~M.; Agarawal,~V.; Gagliardi,~L. Periodic Density
  Matrix Embedding for CO Adsorption on the MgO(001) Surface. \emph{J. Phys.
  Chem. Lett} \textbf{2022}, \emph{13}, 7483--7389\relax
\mciteBstWouldAddEndPuncttrue
\mciteSetBstMidEndSepPunct{\mcitedefaultmidpunct}
{\mcitedefaultendpunct}{\mcitedefaultseppunct}\relax
\EndOfBibitem
\bibitem[Peschel(2012)]{Peschel2012}
Peschel,~I. Special {Review}: {Entanglement} in {Solvable} {Many}-{Particle}
  {Models}. \emph{Braz. J. Phys.} \textbf{2012}, \emph{42}, 267--291\relax
\mciteBstWouldAddEndPuncttrue
\mciteSetBstMidEndSepPunct{\mcitedefaultmidpunct}
{\mcitedefaultendpunct}{\mcitedefaultseppunct}\relax
\EndOfBibitem
\bibitem[Neese(2005)]{Neese2005}
Neese,~F. Efficient and Accurate Approximations to the Molecular Spin-Orbit
  Coupling Operator and Their Use in Molecular g-tensor Calculations. \emph{J.
  Chem. Phys.} \textbf{2005}, \emph{122}, 034107\relax
\mciteBstWouldAddEndPuncttrue
\mciteSetBstMidEndSepPunct{\mcitedefaultmidpunct}
{\mcitedefaultendpunct}{\mcitedefaultseppunct}\relax
\EndOfBibitem
\bibitem[He{\ss} \latin{et~al.}(1996)He{\ss}, Marian, Wahlgren, and
  Gropen]{Hess1996}
He{\ss},~B.~A.; Marian,~C.~M.; Wahlgren,~U.; Gropen,~O. A Mean-field Spin-orbit
  Method Applicable to Correlated Wavefunctions. \emph{Chem. Phys. Lett.}
  \textbf{1996}, \emph{251}, 365--371\relax
\mciteBstWouldAddEndPuncttrue
\mciteSetBstMidEndSepPunct{\mcitedefaultmidpunct}
{\mcitedefaultendpunct}{\mcitedefaultseppunct}\relax
\EndOfBibitem
\bibitem[Bo{\v{c}}a(1999)]{Boca1999}
Bo{\v{c}}a,~R. \emph{Theoretical Foundations of Molecular Magnetism}; Elsevier,
  1999; Chapter 8, pp 371--539\relax
\mciteBstWouldAddEndPuncttrue
\mciteSetBstMidEndSepPunct{\mcitedefaultmidpunct}
{\mcitedefaultendpunct}{\mcitedefaultseppunct}\relax
\EndOfBibitem
\bibitem[Chibotaru(2013)]{Chibotaru2013}
Chibotaru,~L.~F. Ab Initio Methodology for Pseudo-spin Hamiltonians of
  Anisotropic Magnetic Complexes. \emph{Adv. Chem. Phys.} \textbf{2013},
  \emph{153}, 397 -- 519\relax
\mciteBstWouldAddEndPuncttrue
\mciteSetBstMidEndSepPunct{\mcitedefaultmidpunct}
{\mcitedefaultendpunct}{\mcitedefaultseppunct}\relax
\EndOfBibitem
\bibitem[Malrieu \latin{et~al.}(2014)Malrieu, Caballol, Calzado, {de Graaf},
  and Guih{'e}ry]{Malrieu2014}
Malrieu,~J.~P.; Caballol,~R.; Calzado,~C.~J.; {de Graaf},~C.; Guih{'e}ry,~N.
  Magnetic Interactions in Molecules and Highly Correlated Materials:Physical
  Content, Analytical Derivation, and Rigorous Extraction ofMagnetic
  Hamiltonians. \emph{Chem. Rev.} \textbf{2014}, \emph{114}, 429 -- 492\relax
\mciteBstWouldAddEndPuncttrue
\mciteSetBstMidEndSepPunct{\mcitedefaultmidpunct}
{\mcitedefaultendpunct}{\mcitedefaultseppunct}\relax
\EndOfBibitem
\bibitem[Neese and Solomon(2002)Neese, and Solomon]{Neese2002}
Neese,~F.; Solomon,~E.~I. In \emph{Magnetism: Molecules to Materials IV};
  Miller,~J.~S., Drillon,~M., Eds.; Wiley, 2002; Chapter 9, pp 345 --466\relax
\mciteBstWouldAddEndPuncttrue
\mciteSetBstMidEndSepPunct{\mcitedefaultmidpunct}
{\mcitedefaultendpunct}{\mcitedefaultseppunct}\relax
\EndOfBibitem
\bibitem[Maganas \latin{et~al.}(2011)Maganas, Sottini, Kyritsis, Groenen, and
  Neese]{Maganas2011}
Maganas,~D.; Sottini,~S.; Kyritsis,~P.; Groenen,~E. J.~J.; Neese,~F.
  Theoretical {Analysis} of the {Spin} {Hamiltonian} {Parameters} in
  ({CoS4})-{S}-({II}) {Complexes}, {Using} {Density} {Functional} {Theory} and
  {Correlated} Ab Initio {Methods}. \emph{Inorg. Chem.} \textbf{2011},
  \emph{50}, 8741--8754\relax
\mciteBstWouldAddEndPuncttrue
\mciteSetBstMidEndSepPunct{\mcitedefaultmidpunct}
{\mcitedefaultendpunct}{\mcitedefaultseppunct}\relax
\EndOfBibitem
\bibitem[Maurice \latin{et~al.}(2009)Maurice, Bastardis, Graaf, Suaud, Mallah,
  and Guih\'ery]{Maurice2009}
Maurice,~R.; Bastardis,~R.; Graaf,~C.; Suaud,~N.; Mallah,~T.; Guih\'ery,~N.
  Universal theoretical approach to extract anisotropic spin hamiltonians.
  \emph{J. Chem. Theory Comput.} \textbf{2009}, \emph{5}, 2977–2984\relax
\mciteBstWouldAddEndPuncttrue
\mciteSetBstMidEndSepPunct{\mcitedefaultmidpunct}
{\mcitedefaultendpunct}{\mcitedefaultseppunct}\relax
\EndOfBibitem
\bibitem[Andersson \latin{et~al.}(1992)Andersson, Malmqvist, and
  Roos]{Andersson1992}
Andersson,~K.; Malmqvist,~P.; Roos,~B.~O. Second-order perturbation theory
  witha complete active space self-consistentfield reference function. \emph{J.
  Chem. Phys.} \textbf{1992}, \emph{96}, 1218 -- 1226\relax
\mciteBstWouldAddEndPuncttrue
\mciteSetBstMidEndSepPunct{\mcitedefaultmidpunct}
{\mcitedefaultendpunct}{\mcitedefaultseppunct}\relax
\EndOfBibitem
\bibitem[Angeli \latin{et~al.}(2001)Angeli, Cimiraglia, Evangelisti, Leininger,
  and Malrieu]{Angeli2001}
Angeli,~C.; Cimiraglia,~R.; Evangelisti,~S.; Leininger,~T.; Malrieu,~J.-P.
  Introduction of {\emph{n}} -Electron Valence States for Multireference
  Perturbation Theory. \emph{J. Chem. Phys.} \textbf{2001}, \emph{114},
  10252--10264\relax
\mciteBstWouldAddEndPuncttrue
\mciteSetBstMidEndSepPunct{\mcitedefaultmidpunct}
{\mcitedefaultendpunct}{\mcitedefaultseppunct}\relax
\EndOfBibitem
\bibitem[lib()]{liblan_code}
Code is available at {https://github.com/IrisA144/liblan\_preview}.\relax
\mciteBstWouldAddEndPunctfalse
\mciteSetBstMidEndSepPunct{\mcitedefaultmidpunct}
{}{\mcitedefaultseppunct}\relax
\EndOfBibitem
\bibitem[Sun \latin{et~al.}(2018)Sun, Berkelbach, Blunt, Booth, Guo, Li, Liu,
  McClain, Sayfutyarova, Sharma, Wouters, and Chan]{Sun2018}
Sun,~Q.; Berkelbach,~T.~C.; Blunt,~N.~S.; Booth,~G.~H.; Guo,~S.; Li,~Z.~D.;
  Liu,~J.~Z.; McClain,~J.~D.; Sayfutyarova,~E.~R.; Sharma,~S.; Wouters,~S.;
  Chan,~G. K.-L. {PySCF}: The {Python}-based Simulations of Chemistry
  Framework. \emph{Wiley Interdiscip. Rev.: Comput. Mol. Sci.} \textbf{2018},
  \emph{8}, e1340\relax
\mciteBstWouldAddEndPuncttrue
\mciteSetBstMidEndSepPunct{\mcitedefaultmidpunct}
{\mcitedefaultendpunct}{\mcitedefaultseppunct}\relax
\EndOfBibitem
\bibitem[Sun \latin{et~al.}(2020)Sun, Zhang, Banerjee, Bao, Barbry, Blunt,
  Bogdanov, Booth, Chen, Cui, Eriksen, Gao, Guo, Hermann, Hermes, Koh, Koval,
  Lehtola, Li, Liu, Mardirossian, McClain, Motta, Mussard, Pham, Pulkin,
  Purwanto, Robinson, Ronca, Sayfutyarova, Scheurer, Schurkus, Smith, Sun, Sun,
  Upadhyay, Wagner, Wang, White, Whitfield, Williamson, Wouters, Yang, Yu, Zhu,
  Berkelbach, Sharma, Sokolov, and Chan]{Sun2020}
Sun,~Q. \latin{et~al.}  Recent Developments in the {PySCF} Program Package.
  \emph{J. Chem. Phys.} \textbf{2020}, \emph{153}, 024109\relax
\mciteBstWouldAddEndPuncttrue
\mciteSetBstMidEndSepPunct{\mcitedefaultmidpunct}
{\mcitedefaultendpunct}{\mcitedefaultseppunct}\relax
\EndOfBibitem
\bibitem[Sun(2015)]{Sun2015}
Sun,~Q. Libcint: {An} Efficient General Integral Library for {Gaussian} Basis
  Functions. \emph{J. Comput. Chem.} \textbf{2015}, \emph{36}, 1664--1671\relax
\mciteBstWouldAddEndPuncttrue
\mciteSetBstMidEndSepPunct{\mcitedefaultmidpunct}
{\mcitedefaultendpunct}{\mcitedefaultseppunct}\relax
\EndOfBibitem
\bibitem[Schlegel and McDouall(1991)Schlegel, and McDouall]{Schlegel1991}
Schlegel,~H.~B.; McDouall,~J. J.~W. In \emph{Computational Advances in Organic
  Chemistry: Molecular Structure and Reactivity}; {\"O}gretir,~C.,
  Csizmadia,~I.~G., Eds.; Kluwer Academic Publishers, 1991; pp 167 -- 185\relax
\mciteBstWouldAddEndPuncttrue
\mciteSetBstMidEndSepPunct{\mcitedefaultmidpunct}
{\mcitedefaultendpunct}{\mcitedefaultseppunct}\relax
\EndOfBibitem
\bibitem[Pulay(1982)]{Pulay1982}
Pulay,~P. Improved SCF Convergence Acceleration. \emph{J. Comput. Chem.}
  \textbf{1982}, \emph{3}, 556 -- 560\relax
\mciteBstWouldAddEndPuncttrue
\mciteSetBstMidEndSepPunct{\mcitedefaultmidpunct}
{\mcitedefaultendpunct}{\mcitedefaultseppunct}\relax
\EndOfBibitem
\bibitem[Helgaker \latin{et~al.}(2000)Helgaker, J{\o}rgensen, and
  Olsen]{Helgaker2000}
Helgaker,~T.; J{\o}rgensen,~P.; Olsen,~J. \emph{Molecular Electronic-Structure
  Theory}; John Wiley \& Sons, Ltd, 2000\relax
\mciteBstWouldAddEndPuncttrue
\mciteSetBstMidEndSepPunct{\mcitedefaultmidpunct}
{\mcitedefaultendpunct}{\mcitedefaultseppunct}\relax
\EndOfBibitem
\bibitem[Thom and Head-Gordon(2008)Thom, and Head-Gordon]{Thom2008}
Thom,~A. J.~W.; Head-Gordon,~M. Locating Multiple Self-Consistent Field
  Solutions: An Approach Inspired by Metadynamics. \emph{Phys. Rev. Lett.}
  \textbf{2008}, \emph{101}, 193001\relax
\mciteBstWouldAddEndPuncttrue
\mciteSetBstMidEndSepPunct{\mcitedefaultmidpunct}
{\mcitedefaultendpunct}{\mcitedefaultseppunct}\relax
\EndOfBibitem
\bibitem[Gilbert \latin{et~al.}(2008)Gilbert, Besley, and Gill]{Gilbert2008}
Gilbert,~A. T.~B.; Besley,~N.~A.; Gill,~P. M.~W. Self-Consistent Field
  Calculations of Excited States Using the Maximum Overlap Method (MOM).
  \emph{J. Phys. Chem. A} \textbf{2008}, \emph{112}, 13164--13171\relax
\mciteBstWouldAddEndPuncttrue
\mciteSetBstMidEndSepPunct{\mcitedefaultmidpunct}
{\mcitedefaultendpunct}{\mcitedefaultseppunct}\relax
\EndOfBibitem
\bibitem[Carter-Fenk and Herbert(2020)Carter-Fenk, and Herbert]{CarterFenk2020}
Carter-Fenk,~K.; Herbert,~J.~M. State-Targeted Energy Projection: A Simple and
  Robust Approach to Orbital Relaxation of Non-Aufbau Self-Consistent Field
  Solutions. \emph{J. Chem. Theory Comput.} \textbf{2020}, \emph{16}, 5067 --
  5082\relax
\mciteBstWouldAddEndPuncttrue
\mciteSetBstMidEndSepPunct{\mcitedefaultmidpunct}
{\mcitedefaultendpunct}{\mcitedefaultseppunct}\relax
\EndOfBibitem
\bibitem[Briganti \latin{et~al.}(2021)Briganti, Santanni, Tesi, Totti, Sessoli,
  , and Lunghi]{Briganti2021}
Briganti,~M.; Santanni,~F.; Tesi,~L.; Totti,~F.; Sessoli,~R.; ; Lunghi,~A. A
  complete ab initio view of Orbach and Raman spin--lattice relaxation in a
  dysprosium coordination compound. \emph{J. Am. Chem. Soc.} \textbf{2021},
  \emph{143}, 13633--13645\relax
\mciteBstWouldAddEndPuncttrue
\mciteSetBstMidEndSepPunct{\mcitedefaultmidpunct}
{\mcitedefaultendpunct}{\mcitedefaultseppunct}\relax
\EndOfBibitem
\bibitem[Meng \latin{et~al.}(2016)Meng, Wang, Zhang, Leng, Wang, Chen, and
  Gao]{Meng20162}
Meng,~Y.-S.; Wang,~C.-H.; Zhang,~Y.-Q.; Leng,~X.-B.; Wang,~B.-W.; Chen,~Y.-F.;
  Gao,~S. (boratabenzene)(cyclooctatetraenyl) Lanthanide Complexes: A New Type
  of Organometallic Single-ion Magnet. \emph{Inorg. Chem. Front.}
  \textbf{2016}, \emph{3}, 828--835\relax
\mciteBstWouldAddEndPuncttrue
\mciteSetBstMidEndSepPunct{\mcitedefaultmidpunct}
{\mcitedefaultendpunct}{\mcitedefaultseppunct}\relax
\EndOfBibitem
\bibitem[Goodwin \latin{et~al.}(2017)Goodwin, Ortu, Reta, Chilton, and
  Mills]{Goodwin2017}
Goodwin,~C. A.~P.; Ortu,~F.; Reta,~D.; Chilton,~N.~F.; Mills,~D.~P. Molecular
  Magnetic Hysteresis at 60 Kelvin in Dysprosocenium. \emph{Nature}
  \textbf{2017}, \emph{548}, 439--442\relax
\mciteBstWouldAddEndPuncttrue
\mciteSetBstMidEndSepPunct{\mcitedefaultmidpunct}
{\mcitedefaultendpunct}{\mcitedefaultseppunct}\relax
\EndOfBibitem
\bibitem[Liu and Peng(2009)Liu, and Peng]{Liu2009}
Liu,~W.~J.; Peng,~D.~L. Exact Two-component {Hamiltonians} Revisited. \emph{J.
  Chem. Phys.} \textbf{2009}, \emph{131}, 031104\relax
\mciteBstWouldAddEndPuncttrue
\mciteSetBstMidEndSepPunct{\mcitedefaultmidpunct}
{\mcitedefaultendpunct}{\mcitedefaultseppunct}\relax
\EndOfBibitem
\bibitem[Aquilante \latin{et~al.}(2007)Aquilante, Lindh, and
  Pedersen]{Aquilante2007}
Aquilante,~F.; Lindh,~R.; Pedersen,~T.~B. Unbiased auxiliary basis sets for
  accurate two-electron integral approximations. \emph{J. Chem. Phys.}
  \textbf{2007}, \emph{127}, 114107\relax
\mciteBstWouldAddEndPuncttrue
\mciteSetBstMidEndSepPunct{\mcitedefaultmidpunct}
{\mcitedefaultendpunct}{\mcitedefaultseppunct}\relax
\EndOfBibitem
\bibitem[Widmark \latin{et~al.}(1990)Widmark, Malmqvist, and Roos]{Widmark1990}
Widmark,~P.-O.; Malmqvist,~P.-{\AA}.; Roos,~B.~O. Density matrix averaged
  atomic natural orbital {(ANO)} basis sets for correlated molecular wave
  functions. {I}. {First} row atoms. \emph{Theor. Chim. Acta} \textbf{1990},
  \emph{77}, 291--306\relax
\mciteBstWouldAddEndPuncttrue
\mciteSetBstMidEndSepPunct{\mcitedefaultmidpunct}
{\mcitedefaultendpunct}{\mcitedefaultseppunct}\relax
\EndOfBibitem
\bibitem[Roos \latin{et~al.}(2004)Roos, Lindh, Malmqvist, Veryazov, and
  Widmark]{Roos2004}
Roos,~B.~O.; Lindh,~R.; Malmqvist,~P.-{\AA}.; Veryazov,~V.; Widmark,~P.-O. Main
  Group Atoms and Dimers Studied with a New Relativistic ANO Basis Set.
  \emph{J. Phys. Chem. A} \textbf{2004}, \emph{108}, 2851--2858\relax
\mciteBstWouldAddEndPuncttrue
\mciteSetBstMidEndSepPunct{\mcitedefaultmidpunct}
{\mcitedefaultendpunct}{\mcitedefaultseppunct}\relax
\EndOfBibitem
\bibitem[Roos \latin{et~al.}(2008)Roos, Lindh, Malmqvist, Veryazov, Widmark,
  and Borin]{Roos2008}
Roos,~B.~O.; Lindh,~R.; Malmqvist,~P.-{\AA}.; Veryazov,~V.; Widmark,~P.-O.;
  Borin,~A.~C. New Relativistic Atomic Natural Orbital Basis Sets for
  Lanthanide Atoms with Applications to the Ce Diatom and {LuF}3. \emph{J.
  Phys. Chem. A} \textbf{2008}, \emph{112}, 11431--11435\relax
\mciteBstWouldAddEndPuncttrue
\mciteSetBstMidEndSepPunct{\mcitedefaultmidpunct}
{\mcitedefaultendpunct}{\mcitedefaultseppunct}\relax
\EndOfBibitem
\bibitem[Ungur and Chibotaru(2015)Ungur, and Chibotaru]{Ungur2015}
Ungur,~L.; Chibotaru,~L.~F. In \emph{Lanthanides and Actinides in Molecular
  Magnetism}; Layfield,~R.~A., Murugesu,~M., Eds.; Wiley, 2015; Chapter 6, pp
  153 -- 184\relax
\mciteBstWouldAddEndPuncttrue
\mciteSetBstMidEndSepPunct{\mcitedefaultmidpunct}
{\mcitedefaultendpunct}{\mcitedefaultseppunct}\relax
\EndOfBibitem
\bibitem[Wang \latin{et~al.}(2024)Wang, Luo, and Zheng]{Wang2024}
Wang,~Y.; Luo,~Q.-C.; Zheng,~Y.-Z. Organolanthanide Single-Molecule Magnets
  with Heterocyclic Ligands. \emph{Angew. Chem. Int. Ed.} \textbf{2024},
  \emph{63}, e202407016\relax
\mciteBstWouldAddEndPuncttrue
\mciteSetBstMidEndSepPunct{\mcitedefaultmidpunct}
{\mcitedefaultendpunct}{\mcitedefaultseppunct}\relax
\EndOfBibitem
\bibitem[Wang \latin{et~al.}(2023)Wang, Meng, Jiang, Wang, and Gao]{Wang2023}
Wang,~C.; Meng,~Y.-S.; Jiang,~S.-D.; Wang,~B.-W.; Gao,~S. Approaching the
  uniaxiality of magnetic anisotropy in single-molecule magnets. \emph{Sci.
  China Chem.} \textbf{2023}, \emph{66}, 683 -- 702\relax
\mciteBstWouldAddEndPuncttrue
\mciteSetBstMidEndSepPunct{\mcitedefaultmidpunct}
{\mcitedefaultendpunct}{\mcitedefaultseppunct}\relax
\EndOfBibitem
\bibitem[Feng and Tong(2018)Feng, and Tong]{Feng2018}
Feng,~M.; Tong,~M.-L. Single Ion Magnets from 3d to 5f: Developments and
  Strategies. \emph{Chem. Eur. J.} \textbf{2018}, \emph{24}, 7574--7594\relax
\mciteBstWouldAddEndPuncttrue
\mciteSetBstMidEndSepPunct{\mcitedefaultmidpunct}
{\mcitedefaultendpunct}{\mcitedefaultseppunct}\relax
\EndOfBibitem
\bibitem[Cinchetti \latin{et~al.}(2017)Cinchetti, Dediu, and
  Hueso]{Cinchetti2017}
Cinchetti,~M.; Dediu,~V.~A.; Hueso,~L.~E. Activating the molecular spinterface.
  \emph{Nature Mater.} \textbf{2017}, \emph{16}, 507 -- 515\relax
\mciteBstWouldAddEndPuncttrue
\mciteSetBstMidEndSepPunct{\mcitedefaultmidpunct}
{\mcitedefaultendpunct}{\mcitedefaultseppunct}\relax
\EndOfBibitem
\bibitem[Tubman \latin{et~al.}(2016)Tubman, Lee, Takeshita, Head-Gordon, and
  Whaley]{Tubman2016}
Tubman,~N.~M.; Lee,~J.; Takeshita,~T.~Y.; Head-Gordon,~M.; Whaley,~K.~B. A
  Deterministic Alternative to the Full Configuration Interaction Quantum Monte
  Carlo Method. \emph{J. Chem. Phys.} \textbf{2016}, \emph{145}, 044112\relax
\mciteBstWouldAddEndPuncttrue
\mciteSetBstMidEndSepPunct{\mcitedefaultmidpunct}
{\mcitedefaultendpunct}{\mcitedefaultseppunct}\relax
\EndOfBibitem
\bibitem[Sharma \latin{et~al.}(2017)Sharma, Holmes, Jeanmairet, Alavi, and
  Umrigar]{Sharma2017}
Sharma,~S.; Holmes,~A.~A.; Jeanmairet,~G.; Alavi,~A.; Umrigar,~C.~J.
  Semistochastic Heat-Bath Configuration Interaction Method: Selected
  Configuration Interaction with Semistochastic Perturbation Theory. \emph{J.
  Chem. Theory Comput.} \textbf{2017}, \emph{13}, 1595--1604\relax
\mciteBstWouldAddEndPuncttrue
\mciteSetBstMidEndSepPunct{\mcitedefaultmidpunct}
{\mcitedefaultendpunct}{\mcitedefaultseppunct}\relax
\EndOfBibitem
\bibitem[Liu and Hoffmann(2016)Liu, and Hoffmann]{Liu2016}
Liu,~W.; Hoffmann,~M.~R. iCI: Iterative CI toward Full CI. \emph{J. Chem.
  Theory Comput.} \textbf{2016}, \emph{12}, 1169--1178\relax
\mciteBstWouldAddEndPuncttrue
\mciteSetBstMidEndSepPunct{\mcitedefaultmidpunct}
{\mcitedefaultendpunct}{\mcitedefaultseppunct}\relax
\EndOfBibitem
\bibitem[White(1992)]{White1992}
White,~S.~R. Density-matrix Formulation for Quantum Renormalization-groups.
  \emph{Phys. Rev. Lett.} \textbf{1992}, \emph{69}, 2863--2866\relax
\mciteBstWouldAddEndPuncttrue
\mciteSetBstMidEndSepPunct{\mcitedefaultmidpunct}
{\mcitedefaultendpunct}{\mcitedefaultseppunct}\relax
\EndOfBibitem
\bibitem[Chan and Sharma(2011)Chan, and Sharma]{ChanGKL2011}
Chan,~G. K.-L.; Sharma,~S. The Density Matrix Renormalization Group in Quantum
  Chemistry. \emph{Ann. Rev. Phys. Chem.} \textbf{2011}, \emph{62}, 465 --
  481\relax
\mciteBstWouldAddEndPuncttrue
\mciteSetBstMidEndSepPunct{\mcitedefaultmidpunct}
{\mcitedefaultendpunct}{\mcitedefaultseppunct}\relax
\EndOfBibitem
\end{mcitethebibliography}
\providecommand{\latin}[1]{#1}
\makeatletter
\providecommand{\doi}
  {\begingroup\let\do\@makeother\dospecials
  \catcode`\{=1 \catcode`\}=2 \doi@aux}
\providecommand{\doi@aux}[1]{\endgroup\texttt{#1}}
\makeatother
\providecommand*\mcitethebibliography{\thebibliography}
\csname @ifundefined\endcsname{endmcitethebibliography}
  {\let\endmcitethebibliography\endthebibliography}{}

\end{document}